\documentclass[prl, onecolumn, floatfix, amsmath]{revtex4}

\usepackage{graphicx}
\usepackage{epstopdf}
\expandafter\let\csname equation*\endcsname\relax
\expandafter\let\csname endequation*\endcsname\relax
\usepackage{amsmath}
\usepackage{amssymb}
\usepackage{amsthm}
\usepackage[section]{placeins}
\usepackage[UKenglish]{datetime}
\usepackage[english]{babel}
\usepackage{url}

\begin{document}
\title{Optimal control of light propagation through multiple-scattering media in the presence of noise}

\author{Hasan Y{\i}lmaz}\email{h.yilmaz@utwente.nl}
\affiliation{Complex Photonic Systems (COPS), MESA+ Institute for Nanotechnology, University of Twente, P.O. Box 217, 7500 AE Enschede, The Netherlands}

\author{Willem L. Vos}
\affiliation{Complex Photonic Systems (COPS), MESA+ Institute for Nanotechnology, University of Twente, P.O. Box 217, 7500 AE Enschede, The Netherlands}

\author{Allard P. Mosk}
\affiliation{Complex Photonic Systems (COPS), MESA+ Institute for Nanotechnology, University of Twente, P.O. Box 217, 7500 AE Enschede, The Netherlands}

\begin{abstract}
We study the control of coherent light propagation through multiple-scattering media in the presence of measurement noise. In our experiments, we use a two-step optimization procedure to find the optimal incident wavefront. We conclude that the degree of optimal control of coherent light propagation through a multiple-scattering medium is only determined by the number of photoelectrons detected per single speckle spot. The prediction of our model agrees well with the experimental results. Our results offer opportunities for imaging applications through scattering media such as biological tissue in the shot noise limit.
\end{abstract}\maketitle

\newpage

\section{Introduction}
Spatial inhomogeneities in the refractive index of a material such as paper, white paint or biological tissue cause multiple scattering of light. Light propagates diffusively through such materials, which makes the control of light propagation through these kind of materials impossible with conventional optics. A multiple-scattering medium has for a long time been considered as a barrier to optical propagation. It has been theoretically predicted that a multiple-scattering medium can act as a high-precision optical device such as a thin lens, mirror, polarizer or Fourier analyser \cite{Freund1990}. The first optical lens made of multiple-scattering medium was demonstrated by manipulating the incident light field, which starts a new research topic in optics called wavefront shaping\cite{Vellekoop2007_optlett}.  

Many applications of wavefront shaping have recently been demonstrated in advanced optics, biophotonics, nanotechnology, and biomedical imaging \cite{mosk_2012_naturephoton}. Optical pulse compressors have been realized using wavefront shaping \cite{Aulbach2011_prl, Katz2011_nphot, mccabe2011_naturecomm}. It has been shown that a multiple-scattering medium can be used as a high numerical aperture lens \cite{Vellekoop2010_nphot} that enables sub-100 nm optical resolution \cite{Putten2011_PRL}. Recently, wave plates and spectral filters made of multiple-scattering media have been realized \cite{park2012_optexp, silberberg2012_optlett, park2012_optlett, small2012_optlett}. Fluorescence imaging inside biological tissue has been demonstrated by scanning the optical focus guided by acoustic focus \cite{yang2012_naturecomm, cui_2012_naturephoton}. In addition, a non-invasive imaging technique was reported, in which a fluorescent biological object hidden behind a scattering medium was imaged \cite{bertolotti_nature_2012}.

Transforming a multiple-scattering medium into a high-precision optical device requires a high degree of control of light propagation through the medium. Control over the propagation of light through a multiple-scattering medium is quantified by a figure of merit that is given by the intensity enhancement. The enhancement is equal to $\eta=I_\text{opt}/\langle I_0\rangle$ where $I_\text{opt}$ is the intensity in the target after optimization and $\langle I_0\rangle$ is the ensemble averaged intensity in the target before the optimization \cite{Vellekoop2008_optcomm}.

Many wavefront shaping methods have been reported to focus light through multiple-scattering media \cite{Vellekoop2008_optcomm, gigan2010_physrevlett, yang2008_naturephoton,  yang2010_optexpress, psaltis2010_optexpress, Akbulut2011_optexpr, Cui2011_optlett, piestun2012_optexp,piestun2012highspeed_optexp, bifano2012_optexp}. All of these wavefront shaping methods are essentially based on the measurement of a part of the transmission matrix, which is the complex field response of the medium in the transmission to a set of input field bases. Using the information in the transmission matrix, one can synthesize the optimum field to focus light through the medium. In early experiments it was shown that one row of the transmission matrix gives the information to focus light through to a particular position behind the multiple-scattering medium resulting in an enhancement up to $\eta=1000$ \cite{Vellekoop2007_optlett}. An enhancement $\eta=54$ has been reported by Popoff \emph{et al.} \cite{gigan2010_physrevlett}. In their experiment, the transmission matrix of a multiple-scattering medium was measured and the information of the transmission matrix was used  to create a focus through the medium on any selected position \cite{gigan2010_physrevlett}. Using the transmission matrix approach, the transmission of an image is demonstrated through a multiple-scattering medium \cite{Popoff2010_ncomms}. Cui reported an enhancement $\eta=270$ using a parallel optimization method \cite{Cui2011_optlett}. An enhancement $\eta=454$ has been reported by Conkey \emph{et al.} \cite{piestun2012highspeed_optexp}. Park \emph{et al.} reported an enhancement $\eta=400$ \cite{park2012_optexp}. In addition, genetic algorithms have been used for wavefront shaping, and they appear to produce enhancement values in the same range \cite{piestun2012_optexp, silberberg2012_optlett}, suggesting that they are limited by same effects. It has been shown that the enhancement depends linearly on the number of degrees of freedom until it reaches a saturation where practical limitations become prominent \cite{Vellekoop2007_optlett}. However, it remains an open question what is the cause of the wide variation of the saturation value of the enhancement in different experiments.

It has been suggested that measurement noise causes phase errors in the optimization which limit the enhancement \cite{Vellekoop2008_optcomm}. To our knowledge the effect of measurement noise on the enhancement factor has not been investigated. Therefore we present in this paper an experimental and theoretical study of the influence of the noise on the enhancement factor. We show an optimization algorithm that leads to the optimal enhancement in the presence of noise. The optimal enhancement is found to be given by basic physical principles, namely quantum noise in the photodetection process.

\section{The experimental setup}
\begin{figure*}[ht]
\centering
\includegraphics[width=0.6\textwidth]{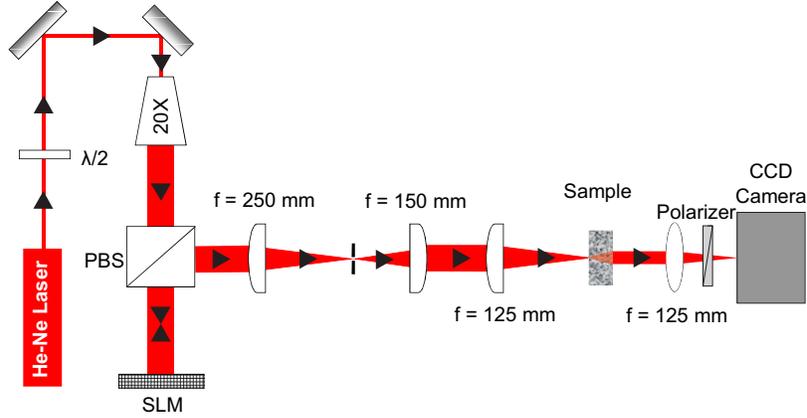}
\caption{The experimental setup for wavefront shaping. Laser light reflected by the SLM is focused on a white ZnO sample. The light transmitted through the sample is detected by a CCD camera. Abbreviations used, SLM: Spatial light modulator, PBS: polarizing beam splitter, $\lambda /2$: half-wave plate, CCD:charge coupled device, 20$\times$: 20$\times$ beam expander.}
\label{fig:setup}
\end{figure*}

Our experimental setup is shown in figure \ref{fig:setup}. A 5 mW He-Ne laser with wavelength $\lambda=632.8$ nm with a noise level of 0.2\% and a long term power drift of 6\% used as the light source. We intentionally use a laser with a high noise and drift. A half wave plate sets the polarization and the beam is expanded to a diameter of 20 mm by a beam expander. The light is transmitted through a polarizing beam splitter and illuminates a spatial light modulator (Holoeye LC-R 2500). The spatial light modulator (SLM) consists of a twisted nematic liquid crystal cell which couples phase and polarization modulation. We used a multipixel modulation method described in reference \cite{Putten2008} to obtain independent phase and amplitude modulation with a single SLM. The two lenses and the pinhole after the polarizing beam splitter are a spatial filter used for the amplitude and phase modulation method. The modulated light is reflected by the polarizing beam splitter and focused on the scattering layer of the sample by a lens with a focal length of 125 mm. The sample is made by spray coating of ZnO nanoparticles on a glass cover slide with a thickness of 170 $\mathrm{\mu m}$. The scattering ZnO layer has a mean free path of 0.65 $\mathrm{\mu m}$ and a thickness of 10 $\mathrm{\mu m}$. The Fourier plane of the backside of the sample is imaged on a CCD camera (Allied Vision Technologies Dolphin F-145B) by a lens with a focal length of 125 mm. A polarizer before the CCD camera selects a single polarization. The CCD signal is read out in counts, where we determined that 1 count corresponds to 1.7 photoelectrons. The CCD camera has a read-out noise  with a variance of 10 $\mathrm{counts}^2$.

\section{The enhancement factor in the presence of noise}
\begin{figure*}[ht]
\centering
\includegraphics[width=0.6\textwidth]{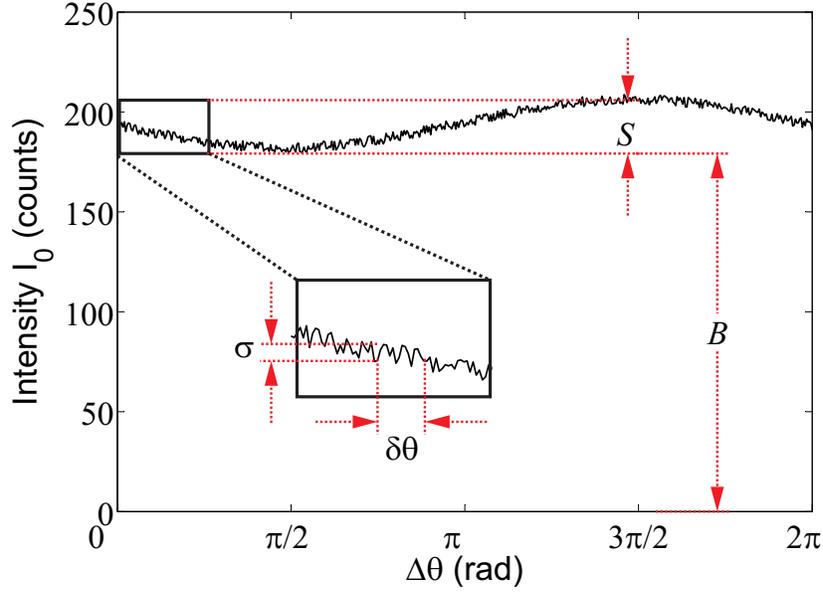}
\caption{The target intensity $I_0$ versus the phase $\Delta\theta$. The modulation signal \emph{S} and the background \emph{B} at the target position are shown during the phase modulation $\Delta \theta$ of a single segment. The standard deviation of the noise is represented by $\sigma$, and the standard deviation of the phase is represented by $\delta\theta$.}
\label{fig:signal}
\end{figure*}

In the wavefront shaping experiment, the SLM surface is divided into a large number of segments $N$. Each segment contains several pixels. The first selected segment is phase modulated between $\Delta\theta=0$ and $\Delta\theta=\mathrm{2\pi}$. We monitor the target signal $I_0$ by integrating the intensity in a disk shaped target area on the CCD with the same size of a single speckle spot while modulating the phase, which results in a sinusoidal signal on top of a background as shown in figure \ref{fig:signal}. We find the phase for the maximum target signal for the corresponding transmission matrix element. The same procedure is applied to all $N$ segments one by one, which yields one row of the transmission matrix. We see in figure \ref{fig:signal} that the target signal on the CCD camera during the phase modulation of a single segment in the presence of noise is

\begin{align}
I_0&=B+S\;\mathrm{cos}\Delta(\theta+\phi)+\sigma,
\label{eq:targetintensity}
\end{align}
where $B$ is the background, $S$ the modulation of the signal, $\Delta\theta$ the phase, $\phi$ the phase offset, and $\sigma$ the standard deviation of noise. The inevitable presence of noise leads to a phase error $\delta\theta$ in the measurement equal to 

\begin{align}
\langle\delta\theta\rangle_\text{RMS}=\frac{\sigma}{S}.
\label{eq:phaseerror} 
\end{align}
Assuming uncorrelated phase errors, the enhancement factor $\eta$ becomes
\begin{align}
\eta& = \frac{\pi}{4} N\langle\mathrm{cos}^2\delta\theta\rangle,
\label{eq:enhancement1}
\end{align}
which is valid for a large number of segments $N>>1$. For small phase errors $(\delta\theta<<1)$ the expression simplifies to

\begin{align}
\eta&= \frac{\pi}{4}N(1-\langle\delta\theta^2\rangle).
\label{eq:enhancement2}
\end{align}
Inserting equation (\ref{eq:phaseerror}) into equation (\ref{eq:enhancement2}) we obtain the enhancement

\begin{align}
\eta&= \frac{\pi}{4}N\bigg(1- \frac{\sigma^2}{S^2}\bigg).
\label{eq:enhancement3}
\end{align}
Equation (\ref{eq:enhancement3}) shows that the enhancement depends both on the number of segments $N$ and also on the noise $\sigma$. Note that the equation (\ref{eq:enhancement3}) is valid under the condition that the modulation signal $S$ is larger than noise $\sigma$.

The modulation signal $S$ depends on the number of segments $N$ as $S\propto N^\text{-1/2}$, whereas the noise $\sigma$ does not depend on $N$. Therefore it is useful to define a normalized signal to noise ratio $R$ that does not depend on the number of segments $N$ as

\begin{align}
R& =\frac{SN^\text{1/2}}{\sigma}.
\label{eq:signaltonoise}
\end{align}
By solving equation (\ref{eq:enhancement3}) for $\sigma$ and inserting the result in equation (\ref{eq:signaltonoise}) we arrive at an expression for the enhancement $\eta$ in which the dependence on $N$ is explicit,

\begin{align}
\eta&= \frac{\pi}{4}N\bigg(1- \frac{N}{R^2}\bigg).
\label{eq:enhancement4}
\end{align}
It is remarkable in equation (\ref{eq:enhancement4}) that the enhancement is not linearly proportional to the number of segments $N$. The enhancement follows a parabolic function which has a maximum. In equation (\ref{eq:enhancement4}), the maximum enhancement is equal to

\begin{align}
\eta_\text{max}&= \frac{\pi R^2}{16}.
\label{eq:enhancementopt1} 
\end{align}
The maximum is obtained by selecting the optimal number of segments to be equal to $N_\text{opt}=R^2/2$. The only way to further increase the enhancement above this maximum is of course to improve the normalized signal to noise ratio $R$.

\section{Pre-optimization}

\begin{figure*}[ht]
\centering
\includegraphics[width=0.6\textwidth]{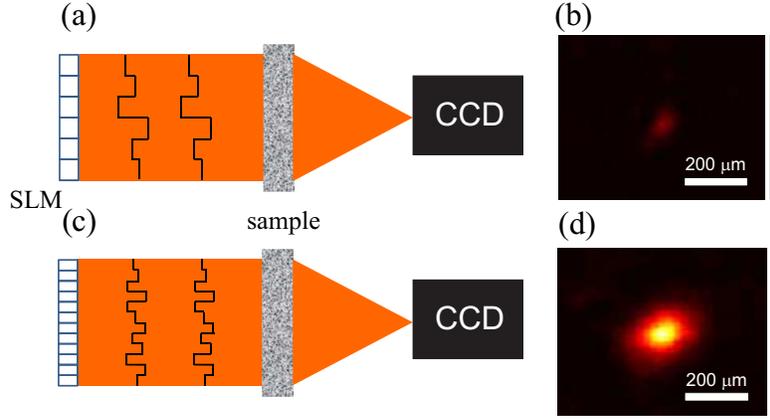}
\caption{The phase map on the SLM after the pre-optimization (a), and after the second optimization (c). The focal spot at the target position after a pre-optimization (b), and after the second optimization (d).}
\label{fig:method}
\end{figure*}

In order to improve the normalized signal to noise ratio $R$, we perform a two-step optimization procedure \cite{Vellekoop2007_optlett}. In figure \ref{fig:method}, we show a schematic of this two-step optimization method. We first perform an optimization with a small number of segments $N_\text{pre}$, leading to an enhancement factor $\eta_\text{pre}$ and a moderate speckle spot (figure \ref{eq:enhancement4}b). As a result of pre-optimization, we obtain a higher modulation signal $S$ on the target speckle spot in the second step see figure \ref{eq:enhancement4}d. In addition, the pre-optimization step provides a locally constant beam profile on the target position, thereby making the measurement robust against mechanical vibrations in the second step. In the second optimization step, we use a much larger number of segments while the previous phase map is retained on the SLM. This means that we continue the second step while we already have focused light on the target position. After the second step, we obtain the final enhancement. We performed the same procedure with different values of $N_\text{pre}$ several times for different values of $\eta_\text{pre}$.

The modulation signal $S$ is an interference signal of the field coming from the modulated single SLM segment with the field coming from the total unmodulated SLM segments. The pre-enhancement $\eta_\text{pre}$ is defined as an intensity enhancement factor. The pre-optimization step only increases the field contribution coming from the total unmodulated SLM segments. Therefore, we conclude that the modulation signal is proportional to the square root of the pre-enhancement factor.

\begin{figure*}[ht]
\centering
\includegraphics[width=0.6\textwidth]{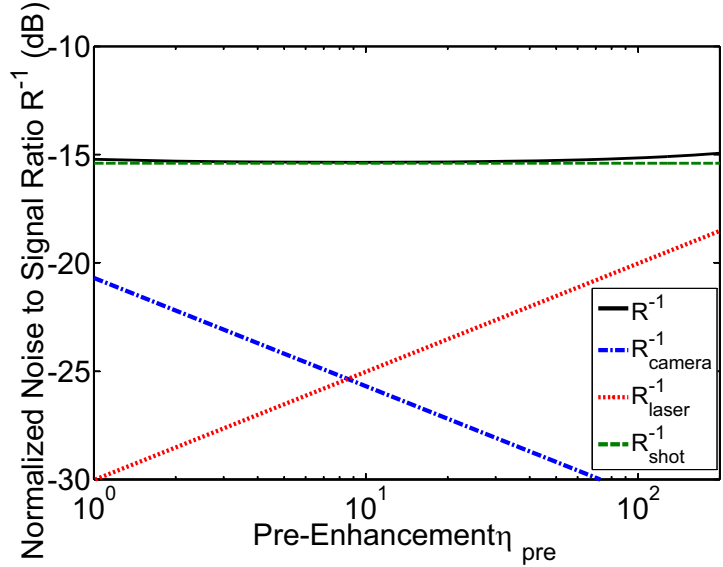}

\caption{The evaluation of three different normalized noise to signal ratio values with the pre-enhancement factor. The black curve shows the total noise to signal ratio. The dashed green curve represents the noise to signal ratio when there is only shot noise, the dashed blue curve the noise to signal ratio when there is only camera read-out noise, and the dashed red curve the noise to signal ratio when there is only laser excess noise.}
\label{fig:noisetosignal}
\end{figure*}

The accuracy of a phase measurement is typically limited by the shot noise and the read-out noise of the camera in phase imaging techniques \cite{mertz_2012}. A wavefront shaping experiment without a pre-optimization step is also limited by the shot noise and the camera read-out noise. The desired effect of the pre-optimization is to increase the modulation signal $S$, but it also has an effect on the noise $\sigma$. The noise depends on the pre-enhancement $\eta_\text{pre}$ in a different way. In our two-step optimization, there are three different significant noise contributions which are (1) the camera read-out noise, (2) the shot noise, and (3) the laser excess noise. In figure \ref{fig:noisetosignal} we show the three types of normalized noise to signal ratio versus $\eta_\text{pre}$ for our experimental situation. The camera read-out noise is suppressed with  a higher $\eta_\text{pre}$. The pre-optimization step improves $R$ when the experiment is limited by the camera read-out noise. The effect of shot noise on $R$ is independent of the pre-optimization step. A higher $\eta_\text{pre}$ leads to a higher the intensity in the target which induces an increase in the intensity fluctuation caused by laser excess noise. Therefore the laser excess noise becomes more dominant with  higher $\eta_\text{pre}$. As a result, an optimal pre-optimization step is useful to suppress the camera read-out noise, which is a dominant noise contribution in a typical wavefront shaping experiment.

We obtain the noise parameters from independent measurements. The noise that arises from the camera read out does not depend on the number of counts on the detector and is simply equal to the variance of the dark counts of the CCD. The standard deviation of shot noise is equal to the square root of the ensemble averaged intensity on the target position measured in number of photoelectrons. The laser excess noise is found by measuring the laser intensity on the target position in time.

\begin{figure*}[ht]
\centering
\includegraphics[width=0.6\textwidth]{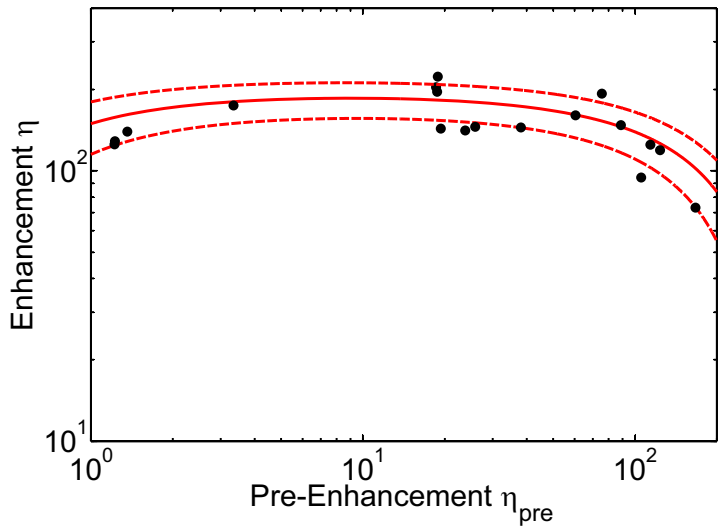}
\caption{The final enhancement versus the pre-enhancement. The black dots show the experimental data. The red curve shows the enhancement according to equation (\ref{eq:enhancement4}) without adjustable parameters. The area between dashed red lines expresses the uncertainty region of the enhancement factor due to intensity drift at the target position during optimization.}
\label{fig:result}
\end{figure*}

In figure \ref{fig:result} we show the measured final enhancement as well as the result of equation (\ref{eq:enhancement4}) versus the pre-enhancement. At low $\eta_\text{pre}$, we observe that the final enhancement rises slightly with $\eta_\text{pre}$, until it reaches a plateau at $\eta_\text{pre}$ is about 10. This rise is due to suppression of the camera read-out noise as more signal impinges on the camera. In the plateau the final enhancement is limited only by the shot noise. A further rise in $\eta_\text{pre}$ decreases the final enhancement due to increase of the laser excess noise. In this case, the enhancement is limited by laser excess noise. The measured enhancements vary with an RMS variation of 60 which is caused by the long term laser power drift. The average laser power during the optimization varies by 6\%. This leads to a change of the shot noise which results in a variation of the enhancement factor. It is seen in figure \ref{fig:result} that the measured enhancement agrees very well with the model predictions with no adjustable parameters.

The pre-optimization step with a $\eta_\text{pre}\approx 10$ suppresses the camera read-out noise which brings the experiment into the shot noise limited regime. In the shot noise limited regime, we obtain the maximum enhancement

\begin{align}
\eta_\text{max}&= \frac{\pi \langle I_0\rangle}{4},
\label{eq:enhancementopt2}
\end{align}
which shows $\eta_\text{max}$ is only proportional to the number of ensemble averaged photoelectrons detected per speckle spot.

\section{Conclusion}

Wavefront shaping experiments reported in literature show a range of enhancements between 50 and 1000. In most of those experiments the limiting factor is likely to be noise. Therefore we have investigated wavefont shaping by feedback in the presence of experimental noise. We distinguish the effect of three types of noise namely the camera read-out noise, the shot noise, and the laser excess noise. Direct optimization is typically limited by the camera read-out noise, which can be reduced using a two-step optimization procedure. Two-step optimization is remarkably robust; even with a low-end camera and a very noisy laser, we show this procedure obtains shot noise limited performance. We obtain a maximum enhancement that is only proportional to the number of photoelectrons detected per single speckle spot.

A wavefront shaping experiment requires the measurement of one row of the transmission matrix of the multiple-scattering medium. A focusing experiment with high enhancement factor is a signature of a precise transmission matrix measurement. Our measurements show that a wavefront shaping experiment is limited by basic physical principles, namely quantized detection of light.  Therefore we conclude that our two-step optimization method can be used to realize shot noise limited transmission matrix measurements. In addition, our method can be used to achieve shot noise limited signal for applications such as imaging through opaque biological tissue.

\section*{Acknowledgements}

We thank Duygu Akbulut, Jacopo Bertolotti, Sebastianus A. Goorden, Pepijn W.H. Pinkse and Elbert G. van Putten  for discussions and Yuwei Chang for participating in the initial measurements. This work is part of the research program of the "Stichting voor Fundamenteel Onderzoek der Materie (FOM)" and the Dutch Technology Foundation (STW), which is financially supported by the "Nederlandse Organisatie voor Wetenschappelijk Onderzoek (NWO)". A.P.M. acknowledges European Research Council grant no. 279248.


\end{document}